\documentclass[twocolumn,prd,groupedaddress,nofootinbib,showpacs]{revtex4}
\usepackage{amssymb}
\usepackage{latexsym}
\usepackage{graphicx}
\usepackage{dcolumn}
\usepackage{bm}
\usepackage{color}

\input{epsf}

\begin{document}

\title{Cosmic string configuration in a  five dimensional
 Brans-Dicke theory }
\author{V. B. Bezerra$^{1a}$ C. N. Ferreira$^{2b}$ and G. de A.
Marques$^{3c}$ }
\affiliation{\\
$^{1}$ Departamento de F\'{\i}sica, Universidade Federal da Para\'{\i}ba,
58059-970, Jo\~ao Pessoa, PB, Brazil \\
\\
$^{2}$ Instituto Federal de Educa\c{c}\~ao, Ci\^encia e Tecnologia Fluminense,\\
28030-130, Campos dos Goytacazes, RJ, Brazil \\
\\
$^{3}$ Unidade Acad\^emica de F\'{\i}sica, Universidade Federal de Campina
Grande,\\
58059-970, Campina Grande, PB, Brazil \\
}
\date{\today}

\begin{abstract}
We consider a scalar field interaction with a cosmic 
string configuration in a flat 3-brane. The origin of this field is given by a
compactification mechanism  in the context of a  5-dimensional 
Brans-Dicke theory.  We use the Cosmic Microwave Background Radiation constraint to 
analyse a possibility of optical activity effect in connection with the Brans-Dicke 
parameter $\omega$ and  obtain relations between the compactification modes. We show that 
the dilatons produced by a cosmic string can  decay into gauge bosons with masses given 
by the compactification modes. The Brans- Dicke parameter $\omega  $ imposes stringent constraints 
on the mass of the dilaton which lives in the brane and help us to understand the energy scales. In
this scenario we estimate the  lifetime of the dilaton which decays into light gauge bosons and the 
dependence of  this phenomenon with the Brans-Dicke parameter.
\end{abstract}

\maketitle

PACS numbers{04.20-q, 04.50+h}

\section{Introduction}

Cosmic strings\cite{Vilenkin1} could have been produced in the early stages of the universe 
as a consequence of cosmological phase transitions \cite{Kibble80}. This string may produce interesting effects associated with the conical topology of its spacetime.
As examples of these effects we can mention the emission of radiation by a freely moving body \cite{Audretsch}
and the vacuum fluctuations of quantum fields \cite{Heliwell86}, among others.
It has been argued that gravitational interaction may be described by scalar-tensor 
theory, at least at sufficiently high energy scales \cite{Green87}. 
Thus it seens natural to investigate some physical systems, as for
example, one which contains scalar fields and  cosmic strings, at these energy 
scales. In particular, it is interesting to consider this system in the framework of the 
Brans-Dicke theory of gravity \cite{Brans}.

At high energy scales, usually, the theories of fundamental physics are formulated 
in higher dimensional spacetimes in which  case  is assumed that the extra dimensions are compactified. 
This compactification of spatial dimensions leads to interesting quantum effects like as topological mass
generation\cite{Ford79} and symmetry breaking \cite{Odnitsov88}.

In despite of the successes of the Standard Model of elementary particles,
the fact that the gravitational interaction is not included and the
hierarchy problem is not considered constitute unsatisfactories
aspects of this model. In the context of a String Theory which has been considered as
a promise theory, in which these questions can be explained. Moreover, based on
String Theory ideas, the Brane-World scenario has been proposed to take into
account the hierarchy between the electroweak and Planck scales and
becomes the effective form to realize the cosmological scenarios. 

On the other hand, the consistency of String Theory, which is a candidate to
describe all fundamental interactions, requires that our world has to have more than
four dimensions. Originally, extra dimensions were supposed to be very
small (of the order of Planck length). However, it has been proposed, recently, that
the solution to the hierarchy problem may arises by considering some of
these extra dimensions to be not so small. In the approach used in\cite{ADD} the
space has one or more flat extra dimensions while, according to the so
called Randall-Sundrum (RS) model\cite{RS}, hierarchy would be explained by
one large warped extra dimension. In the RS approach the 4D world is a
3-brane embedded in a 5-dimensional anti-de Sitter (AdS) bulk. In this 
framework, the Standard
Model fields are confined to the 3-brane while gravity propagates in the
bulk. In this scenario the hierarchy is generated by an exponential function
of the compactification radius, called warp factor.

The Randall-Sundrum model is defined in a 5-dimensional AdS slice
characterized by a background metric written as

\begin{equation}
ds^2 = e^{2 \sigma ( \Omega ) } \eta_{\mu \nu} dx^{\mu} dx^{\nu } + 
d\Omega^2  \label{1}
\end{equation}

\noindent where $x^{\mu}$ are Lorentz coordinates on the four-dimensional
surfaces of constant $\Omega$, which takes values in the range 
$-L \leq \Omega \leq L$, with (x, $
\Omega$) and (x, $ - \Omega$) identified, and the 3-brane located at $\Omega =
0, L$. Here, $L$ sets the size of the extra dimension, $\sigma(\Omega) = -\kappa
L |\Omega |$ and k is taken to be of the order of the Planck scale. In this approach,
gravity propagates in the bulk while in the Standard Model fields propagate on
the D-3 brane defined at $\Omega = L$. The energy scales are related in
such a way that the TeV mass scales are produced on the brane from Planck masses
through the warp factor $e^{-2\kappa L}$

\begin{eqnarray}
m & = & m_0 \,e^{-2\kappa L} \\
M_p^2 & = & {\frac{M^3 }{\kappa}} \Big( 1 - e^{-2 \kappa L}\big)
\label{hierarchy1}
\end{eqnarray}

\noindent In our case we will deal with massless fields in the bulk but this
warp factor will appear in the effective couplings in four dimensions.

This paper is organized as follows. In section II we discuss the compactification
of the scalar field. In section III the charged cosmic string configuration 
in the brane world scenario is presented. In sec IV we discuss the cosmic 
optical activity and its connection with the Cosmic Microwave Background Radiation. In section V, 
we consider particle production in the Brane-World scenario. Finally, in section VI, we present the conclusions.

\section{Compactification of the scalar field perturbation in the Brane-World scenario}

In this section we consider a model that represents the matter fields in
the 3-brane with a five dimensional scalar-tensor theory and study the
compactification of the scalar field perturbation in the spacetime of the
scalar-tensor Brane-World scenario. The complete action that represents the system
under consideration can be written as 
\begin{eqnarray}
\mathcal{S}\! \! \! &=& \! \! \!\! \int d^5x \, \sqrt{-\tilde g}\, \tilde
\Psi \, \, \Big[R - \Lambda \, + \, {\frac{\omega }{\tilde \Psi^2 }}\, \, 
\tilde{ g}^{\hat \mu \hat \nu}\, \partial_{\hat \mu}\, \tilde \Psi
\partial_{\hat \nu} \tilde \Psi \Big]  \nonumber \\
& -& \! \! \!\! \int_{_{\Omega =0}} \! \! d^{4}x \, \sqrt{- \tilde h}\, \, \tilde
\Psi \,\lambda_1 + \int_{_{\Omega =L}} \! \! d^{4}x \, \sqrt{- \tilde h}\, \, \tilde
\Psi \,\lambda_2 \, ,  \label{acao1}
\end{eqnarray}

\noindent where the label $\hat{\mu}=(\mu ,\Omega )$, with $\mu =0...3$ and
the compact five dimensional coordinate $\Omega $ taking values in the range 
$-L\leq \Omega \leq L$,
with the identification $\Omega =-\Omega $. We consider $\tilde{h}_{\mu \nu }
$ as the induced metric on the two branes, $ \Omega =0$ and $\Omega =L $, 
$\Lambda $, a potential in the bulk and 
$\lambda _{1, 2 }$ the brane potential. The constant $\omega $ is the
Brans-Dicke parameter. The orbifold condition implies that the solution $%
\tilde{\Psi}$ and the metric are functions of $\Omega $ and $
x$ being odd function with respect to $ x $.

The metric with matter on the brane can be written as

\begin{equation}
ds^2 = \bar g_{\hat \mu \hat \nu}dx^{\hat \mu} dx^{\hat \nu} =
e^{2\sigma}g_{\mu \nu}dx^{\mu \nu} +d\Omega^2
\end{equation}
where $g_{\mu \nu}$ is the metric on the brane and depends on $x$-cordinate.

We consider the background equations in action (\ref{acao1}) and with the
BPS solutions we find that the scalar field $\tilde \Psi$ and the potentials
are given by 
\begin{eqnarray}
\tilde \Psi &= & Ce^{{\frac{\sigma }{\omega +1}}}, \\
\lambda_{1,2} & = & \pm \lambda \tilde \Psi . \\
\end{eqnarray}

The relation between the potentials $\lambda $ and $\Lambda$ are 
\begin{equation}
\lambda = 4\left({\frac{3\omega +4 }{4 \omega +5}}\right)^{1/2}\sqrt{%
-\Lambda}.
\end{equation}

We can study the hierarchy problem in this framework considering that our
brane is localized at $\Omega = L $ and that $\tilde g_{\mu \nu}(x,\Omega) =
e^{2 \sigma} g_{\mu \nu}(x)$, $\tilde \Psi( x, \Omega) = \tilde \Psi(\Omega) 
$, $\tilde g_{\Omega \Omega}(x, \Omega) = 1 $ and $\tilde g_{\mu \Omega}(x,
\Omega) = 0$. Thus,  from action (\ref{acao1}) we have

\begin{equation}
S_{Grav}= \int_{-L}^{L} d\, \Omega \, \tilde \Psi \, e^{4 \sigma}\int d^4x
\, \, \sqrt{-g}\, \,\, \, R_{4}
\end{equation}
where $R_4$ is the Ricci scalar in four dimensions.

We can perform the $\Omega $ integral to obtain the following four-dimensional action

\begin{equation}
S_{eff}^{Grav} = \frac{1}{16 \pi G_{eff}}\int d^4x\sqrt{-g} R_{4}\label{actiongrav}
\end{equation}

After the integration in the fifth coordinate, we have

\begin{equation}
2 M_{Pl}^2 = \frac{2 C (\omega +1)}{4 \omega + 5}{\frac{1 }{k }} \left( 1 -
e^{- {\frac{4 \omega + 5 }{\omega +1}}k L} \right)
\end{equation}
where $2 M_{Pl}^2 = {\frac{1 }{16 \pi G_{eff}}} $ and we are considering

\begin{equation}
k  = (\omega +1) \sqrt{{\frac{- \Lambda }{(3 \omega + 4)(4 \omega +5)}}}%
\end{equation}

\begin{equation}
C = \left({\frac{1 }{16 \pi G_{eff}}}\right)^{3/2}
\end{equation}

It is worth calling attention to the fact that this choice is compatible with eq. (\ref{hierarchy1}) 
and has dimension of $M^{3}$.

Now, we study the compactification of the scalar field fluctuation in this
scenario. We consider the action that represent this fluctuation as given
by

\begin{equation}
\mathcal{S}^{5}_{scalar}=-\int d^{5}x\sqrt{-g_{5}}\,\tilde{\Psi}\partial _{%
\hat{\mu}}\Phi \partial ^{\hat{\mu}}\Phi ,  \label{acao2}
\end{equation}%
which is compatible with the scalar action in the  Einstein frame. Thus the complete action turns into

\begin{eqnarray}
S_{_{scalar}} &=&- C \int d^5x\sqrt{-g} \Big[ e^{-{\frac{4 \omega +5 }{\omega
+1}} k \Omega}g^{\mu \nu} \partial_{\mu} \Phi \partial_{\nu} \Phi  \nonumber
\\
& & + \Phi \partial_{\Omega}\Big( e^{-{\frac{4 \omega +5 }{\omega +1}} k
\Omega}\partial_{\Omega} \Phi\Big)\Big]
\end{eqnarray}
where $\Phi $ is a function of the five-dimensional coordinate. 
Now, let  write the decomposition of the scalar field $\Phi(x, \Omega)$ as a
sum over modes as follows:

\begin{equation}
\Phi(x, \Omega) = {\frac{1 }{\sqrt{L}}}\sum_n X_n(x) \xi_n(\Omega) ,
\label{expan1}
\end{equation}
where the modes $\xi_n(\Omega)$ satisfy

\begin{equation}
{\frac{1 }{L}}\int_{-L}^{L} d \Omega e^{{\frac{2 \omega +3 }{\omega +1}}
\sigma} \xi_n(\Omega) \xi_m(\Omega)= \delta_{nm}
\end{equation}

\noindent and

\begin{equation}
-{\frac{e^{2\sigma}}{v(\Omega)}}\frac{d}{d \Omega} \Big( v(\Omega) \frac{d
\xi_n}{d\Omega}\Big)= m_n^2 \xi_n, \label{besse0}
\end{equation}
with $v(\Omega) = e^{2 \nu \sigma} $ and $\nu = {\frac{4
\omega +5 }{2(\omega +1)}}$. After the compactification we have

\begin{equation}
S_{_{X_n}} =C \sum_n \int d^4 x \sqrt{-g}\Big(\partial_{\mu}X_n \partial^{\mu}X_n
+m_n^2 X_n^2\Big).\label{actionscalar}
\end{equation}

The zero mode occurs when $m_{n}=0$ and give us the solution

\begin{equation}
\xi_0 = {\frac{1 }{N_0}}e^{-2\nu \sigma} \label{zeromode}
\end{equation}
where the normalization constant is given by

\begin{equation}
N_0 = {\frac{1 }{\sqrt{(\nu +1) k L}}} e^{(\nu +1)k L}
\end{equation}

As in usual Kaluza-Klein compactifications, the bulk field $\Phi(x, \Omega) $
manifests itself to a four dimensional observer as an infinite set
of scalars $X_n(x)$ with masses $m_n$, which can be found  by solving eq. (\ref{besse0}). 
After changing variables to $z = e^{k \Omega}/k$, then, the  equation of motion can be written as

\begin{equation}
z^u{\frac{ d }{d z}}\left(z^{-u}\frac{d\xi_n}{d z}\right) = - m_n^2 \xi_n
\label{Bessel1}
\end{equation}
where $u = {\frac{3 \omega + 4 }{\omega +1}}$. The solutions of this
equation are combination of Bessel functions of order $\nu $ and can be written as

\begin{equation}
\xi_n(\Omega) = {\frac{e^{-\nu \sigma(\Omega)} }{N_n}} \Big[J_{\nu}(x_{n\nu}) + b_{n \nu}Y_{\nu}(x_{n\nu})\Big],  \label{Bessel2}
\end{equation}
where the normalization constant is expressed as

\begin{equation}
N_n={\frac{2 e^{kL}}{x_{n\nu}\sqrt{2kL}}}\left(\int d z z \Big[J_{\nu}(m_nz)
+ b_{n \nu}Y_{\nu}(m_nz)\Big]^2\right)^{1/2},
\end{equation}
with $x_{n\nu}=\frac{m_n }{k}e^{k L}$.

Let us consider the solution continuous on the branes,
which  means that the Neumann condition are valid.  Therefore,
at $\Omega =0$, we have

\begin{equation}
b_{n\nu} = -{\frac{ J_{\nu-1}(x_{n\nu}e^{-k L}) }{Y_{\nu-1}(x_{n\nu}e^{-k
L}) }},
\end{equation}
Applying the same condition at $\Omega = L$, we find

\begin{eqnarray}
x_{n \nu}^2e^{-kL} [J_{\nu -1}(x_{n \nu})Y_{\nu -1}(x_{n \nu}e^{-kL}) 
\nonumber \\
- Y_{\nu -1}(x_{n \nu})J_{\nu -1}(x_{n \nu}e^{-kL}]=0  \label{zeroseq}
\end{eqnarray}

\begin{figure}
\centering
\includegraphics[width=7.5cm]{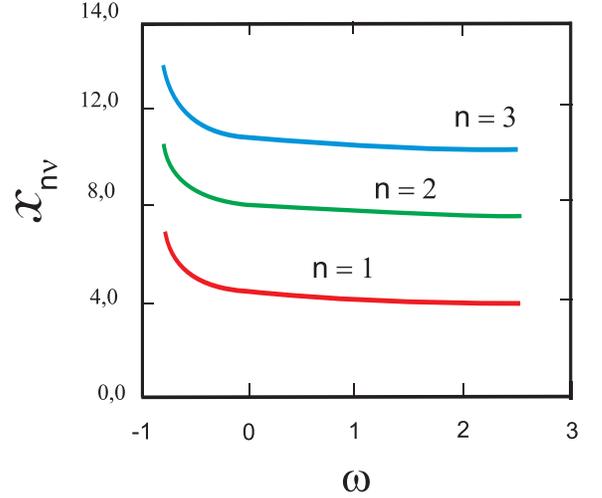} 
{\caption{First mode $ x_{n \nu}$ as a function 
\\of $\omega $. }\label{Figura1}}
\end{figure}

Note that the action  (\ref{actionscalar}) differ from dilaton action in Einstein frame in four dimensions 
only by the presence of n massive modes. In zero mode the mass vanishes and we use it to calculate the zero 
mode function $ \xi_0$ given by Eq. (\ref{zeromode}). In this case the action is similar to the four-dimensional dilaton action. The masses given by the zeros arised from the Neumann conditions in the brane localized in 
$\Omega = L$. The important fact is that the brane conditions and the normalization determine the solution 
of the Bessel equation (\ref{Bessel2}). 
In the Fig. 1 it is showed the dependence of the modes $ x_{n \nu}$ as a function of the Brans-Dicke 
parameter which in next sections will be interpreted as the fine tune paramenter. We can see that this 
dependence presents an asymptotical behavior around $\omega = -1$.

\section{Charged Cosmic String Configuration in the Brane-World Scenario}

In this section we study how this framework affect the cosmic string configuration in the brane. 
By analogy with the four dimensional Brans-Dicke theory we consider similar  dilaton coupling in 
Einstein frame. In this case the action corresponding to a cosmic string can be written as 

\begin{equation}
 {\cal S}_{CS}=\int d^4x \sqrt{\bar g} \, \, {\cal L}(\bar g_{\mu \nu}, \phi, H_{\mu}),
\end{equation}
where 

\begin{eqnarray}
 \bar g_{\mu \nu} &= &e^{2 \alpha \Phi(x, \Omega =L)} g_{\mu \nu} \nonumber\\
 \Phi(x, \Omega =L) &=& {1 \over \sqrt{L}} \sum_n X_n (x)\,\xi_n(\Omega =L). \nonumber
\end{eqnarray} 

Now, let us investigate the cosmic string model assuming that the spacetime is
approximately flat. Thus, the cosmic string action on the brane that
presents a coupling with the dilaton modes at $\Omega = L$ can be written as

\begin{eqnarray}
\mathcal{S}_{mat} & =& \Pi_{\! \! \! \! _{_{n}}}\int d^4x \Big[-\frac{1}{2}
e^{2\alpha \xi_nX_n} D_{\mu}\phi (D^{\mu}\phi)^*  \nonumber \\
& & - \frac{1}{4} H_{\mu \alpha}H^{\mu \alpha} -e^{4\alpha \xi_nX_n}V(\phi)%
\Big] ,  \label{acao3-1}
\end{eqnarray}
where  $\xi_n $ at $(\Omega =L)$ is given by equation (\ref{Bessel2}).

The action given above has a $U(1) $ symmetry, associated with the $\phi$%
-field, which is broken by the vacuum and gives rise to vortices of the
Nielsen-Olesen type \cite{Nielsen}

\begin{equation}
\begin{array}{ll}
\phi = \varphi(r )e^{i\theta - k L}, &  \\ 
H_{\mu} = \frac{1}{q}[P(r) - 1]\delta^{\theta}_{\mu}, & 
\end{array}
\label{vortex1}
\end{equation}

\noindent in which $(t,r,\theta,z)$ are usual cylindrical coordinates. The
boundary conditions for the fields $\varphi(r) $ and $P(r)$ are the same as
those of ordinary cosmic strings\cite{Nielsen}, namely,

\begin{equation}
\begin{array}{ll}
\begin{array}{ll}
\phi(r) = \eta e^{i\theta -k L} & r \rightarrow \infty \\ 
\phi(r) =0 & r = 0%
\end{array}
& 
\begin{array}{ll}
P(r) =0 & r \rightarrow \infty \\ 
P(r) =1 & r= 0,%
\end{array}%
\end{array}
\label{config1}
\end{equation}

\noindent where $D_{\mu} \phi = ( \partial_{\mu} + i qH_{\mu})\phi $ is the
covariant derivative. The potential $V(\phi)$ triggering the spontaneous
symmetry breaking can be fixed by

\begin{equation}
V(\phi) = \frac{\lambda_{\phi}}{4} ( |\phi | ^2 - \tilde \eta^2)^2,
\end{equation}
\noindent where $\tilde \eta = \eta e^{-k L} $ and $\lambda_{\phi}$ is the
coupling constant. Constructed in this way, this potential possesses all the
ingredients that makes it to induce the formation of an ordinary
cosmic string.

Let us consider that the dilaton field, propagating in the bulk is
given by Maxwell-Chern-Simons three-tensor on the brane. After the
integration in $\Omega$-coordinate, we have 

\begin{eqnarray}
\mathcal{S}_{free} & \! \! = & -{\frac{1 }{4}} \int d^4x \sqrt{-{\ g}}F_{\mu
\nu}F^{\mu \nu}  \label{actionint}
\end{eqnarray}

\begin{eqnarray}
\mathcal{S}_{int} & \! \! = & \int d^4x \sqrt{-{\ g}}\Big[\sqrt{(\nu +1)kL}%
\,e^{-(\nu -1) k L} \partial_{\mu} X_0  \nonumber \\
& &\!\! \! \! \! \!\! \! \! \!\!\! \! \! \! \!\! \! \! \! \! \! + \,e^{\nu
kL}\! \!\sum_{n=1} \!{\frac{J_{\nu}(x_{n\nu}) }{N_n}} \! \partial_{\mu}X_n
\! \Big]\Big[\alpha_{_+}^{^{\! \!(1)}} \!H_{\nu}\tilde H^{\mu \nu} \!\! + \!
\alpha_{_+}^{^{\!(2)}}\! \! A_{\nu}\tilde H^{\mu \nu}\Big]
\label{actionint1}
\end{eqnarray}
where $H_{\mu \nu} $ is a field strength associated with a gauge field and  
$\alpha_{_+}^{^{\!(1)}} $, $\alpha_{_+}^{^{\!(2)}} $ are constant coefficients.

The action (\ref{actionint1}) is the general form of the interaction terms,
and is  dictated by  the effects that we want to analyse. The consequences of the 
action (\ref{actionint1}) to the cosmic string configuration can be analyzed by the 
equation of the motion of the gauge fields. If we consider the Ansatz for the gauge field $%
A_{\mu }(r)$ and the equation of motion for the gauge field $H_{\mu }(r)$ in
weak field approximation, we have that the equation of the motion in Minkowiski
space is 

\begin{equation}
\partial_{\mu} H^{\mu \nu} + \epsilon^{\nu \alpha \beta \mu} \sum_n\xi_n%
\Big[ \alpha_+^{(1)} H_{\alpha \beta} + \alpha_+^{(2)}F_{\alpha \beta}\Big]%
\partial_{\mu} X_n =J^{\nu},  \label{H}
\end{equation}
where $j_{\mu}(\rho/\epsilon_0,\mu J_z)$ is given by

\begin{equation}
J_{\mu} = {\frac{-iq }{2}} \Big[\phi^* D_{\mu}\phi - \phi (D_{\mu} \phi)^* %
\Big].
\end{equation}
Note that if the charge $J^0$ is different from zero, then $H_t \neq 0$. The Ansatz compatible with this configuration is

\begin{eqnarray}
H_t(r) =0 & r \rightarrow \infty  \nonumber \\
H_t (r)= b & r=0
\end{eqnarray}

The charge behavior can be analyzed in the weak field approximation of the
dilaton field as

\begin{equation}
A(\Phi) = A(\Phi_{(0)}) + A{\acute{}}
(\Phi_{(0)}) \Phi_{(1)}.
\end{equation}

Considering the equation of motion for the external field $F_{\mu \nu }$, 
we have

\begin{equation}
\partial_{\lambda} F^{\lambda \nu} - \alpha_+^{(2)}\epsilon^{\mu \nu \alpha
\beta} \sum_n \xi_n\partial_{\mu} X_n H_{\alpha \beta}=0 \label{ext}
\end{equation}

Consider the electric field $ E^i$ and magnetic field $B^i$, defined as
$ E^i = F^{0 i}$ and  $B^i = - \epsilon^{ijk}F_{jk} $. Then, we obtain

\begin{eqnarray}
E_{ext} &=&{\frac{1}{2\pi \epsilon r}}\Big[\sqrt{(\nu +1)kL}\,e^{-(\nu
-1)kL}\int_{0}^{r_{0}}\partial _{z}X_{0}B(r)rdr  \nonumber \\
&&e^{\nu kL}\sum_{n=1}{\frac{J_{\nu }(x_{n}\nu )}{N_{n}}}\int_{0}^{r_{0}}%
\partial _{z}X_{n}B(r)rdr\Big]
\end{eqnarray}%
where $\varepsilon =1/\alpha _{+}^{(2)}$ This result show up that the
external field generated by the charged cosmic string in this framework depends of
the brane world parameters.

The energy-momentum tensor $T_{\mu \nu}$ is given by

\begin{eqnarray}
T^t_t & = & - {\frac{1}{2}} A^2\Big[ \varphi^{\prime 2} + {\frac{1 }{r^2}}
\varphi^2P^2+ A^{-2}\left({\frac{A_t^{\prime 2} }{4 \pi e^2}}\right) 
\nonumber \\
&+ & \varphi^2 H_t^2 + A^{-2} {\frac{1}{r^2}} \left( {\frac{P^{\prime 2} }{4
\pi q^2}}\right) + 2 A^2 V(\varphi)  \nonumber \\
& + & A^{-2} \left({\frac{H_t^{\prime 2} }{4 \pi e^2}}\right)\Big]
\end{eqnarray}

\begin{eqnarray}
T^z_z & = & - {\frac{1}{2}} A^2\Big[ \varphi^{\prime 2} + {\frac{1 }{r^2}}
\varphi^2P^2- A^{-2} \left({\frac{A_t^{\prime 2} }{4 \pi e^2}}\right) 
\nonumber \\
&- & \varphi^2 H_t^2+ A^{-2} {\frac{1 }{r^2}} \left( {\frac{P^{\prime 2} }{4
\pi q^2}}\right) + 2 A^2 V(\varphi)  \nonumber \\
& - & A^{-2} \left({\frac{H_t^{\prime 2} }{4 \pi e^2}}\right)\Big]
\end{eqnarray}

\begin{eqnarray}
T^r_r & = & {\frac{1}{2}} A^2\Big[ \varphi^{\prime 2} - {\frac{1 }{r^2}}
\varphi^2P^2- A^{-2}\left({\frac{A_t^{\prime 2} }{4 \pi e^2}}\right) 
\nonumber \\
&+ & \varphi^2 H_t^2 + A^{-2} {\frac{1 }{r^2}} \left( {\frac{P^{\prime 2} }{%
4 \pi q^2}}\right) - 2 A^2 V(\varphi)  \nonumber \\
& - & A^{-2} \left({\frac{H_t^{\prime 2} }{4 \pi e^2}}\right)\Big]
\end{eqnarray}

\begin{eqnarray}
T^{\theta}_{\theta} & = & - {\frac{1}{2}} A^2\Big[ \varphi^{\prime 2} - {%
\frac{1 }{r^2}} \varphi^2P^2- A^{-2} \left({\frac{A_t^{\prime 2} }{4 \pi e^2}%
}\right)  \nonumber \\
& +&\varphi^2 H_t^2 - A^{-2} {\frac{1 }{r^2}} \left( {\frac{P^{\prime 2} }{4
\pi q^2}}\right) + 2 A^2 V(\varphi)  \nonumber \\
& - & A^{-2} \left({\frac{H_t^{\prime 2} }{4 \pi e^2}}\right)\Big]
\end{eqnarray}

In this analysis we consider the weak field approximation, which permits us to expand the
dilaton field as 

\begin{equation}
X_n = X_{(0)n} + \epsilon X_{(1)n}  \label{expansion}
\end{equation}
where $X_0$ is the constant dilaton value in the background without cosmic
string.

In our framework we find the equation for the field $X_n$, which reads as

\begin{eqnarray}
\Box X_{n} &=&-{\frac{1}{4}}{\frac{dU(X_{n})}{dX_{n}}}-4\pi G_{eff}C^{-1}\xi
_{n}\Big[\alpha T  \nonumber \\
&+&2\Big(\alpha _{+}^{(1)}H_{\mu \nu }+\alpha _{+}^{(2)}F_{\mu \nu }\Big)%
\tilde{H}^{\mu \nu }\Big],  \label{dilatonequation}
\end{eqnarray}
where the coefficients $\alpha_{_+}^{^{\!(1)}}  $  and $ \alpha_{_+}^{^{\!(2)}} $ are 
associated with the topological contributions. The terms whose coefficients are $\alpha_{_+}^{^{\!(1)}} $ and  $\alpha_{_+}^{^{\!(2)}}$ vanishe because the fields does not depend on $t$ and $z$. Thus, we obtain  

\begin{equation}
\Box X_n = -{\frac{1}{4}} {\frac{d U(X_n) }{d X_n}} - 4 \pi G_{eff} C^{-1}
\xi_n \alpha T
\end{equation}

Let us consider the solution as

\begin{equation}
X_{n}(t,r,z)=\chi _{n}(r)+f(r)\Xi _{n}(t,z)  \label{X2}
\end{equation}%
where $f(r)$ is required to vanishes outside the string core. The charge
induced by the dilaton is

\begin{eqnarray}
Q\! &\!= \!&\!2\pi\alpha_+^{^{\!(1})}\!\Big[ \! \alpha \sqrt{(\nu \!+ \!1)k L}\, e^{-(\nu
-1) k L} \partial_z\Xi_0 \int_0^{r_0}\! \! \! \! f(r)B(r) r dr  \nonumber \\
& & \,e^{\nu kL}\! \!\sum_{n=1} {\frac{J_{\nu}(x_{n\nu}) }{N_n}}
\partial_z\Xi_n \int_0^{r_0} f(r)B(r) r dr\Big]  \label{charge1}
\end{eqnarray}

Using solution (\ref{X2}), we obtain to first order in $G_{eff}$, the equation

\[
\chi _{n}^{\prime \prime }+{\frac{1}{r}}\chi _{n}^{\prime }+m_{n}^{2}\chi
_{n}=-\,4\pi \,G_{eff}\,\xi _{n}\,C^{-1}\alpha \,\tilde{T}_{(0)}
\]%
The field $\Xi $, obeys the following equations
\begin{eqnarray}
{\frac{\partial ^{2}f}{\partial r^{2}}}+{\frac{1}{r}}{\frac{\partial f}{%
\partial r}} &=&\omega _{n}^{2}f  \label{f} \\
{\frac{\partial ^{2}\Xi _{n}}{\partial t^{2}}}-{\frac{\partial ^{2}\Xi _{n}}{%
\partial z^{2}}} &=&\omega _{n}^{2}\Xi _{n}  \label{rho}
\end{eqnarray}%
where $f$ satisfies the boundary condition at $ r \rightarrow \infty $.
The arbitrary constant $w_{n}$ can assume, a priori, both positive and
negative values for each n as $\omega ^{2}=k^{2}-\omega _{0}^{2}$. The
solution of the equations of  motion (\ref{f}) and (\ref{rho}) are
important to discuss the aspects connected with the change the charge. Considering the well known 
electric case, let us take $w=k$. Thus, the solution for $f$ is

\begin{equation}
f(r)=f_{I}I_{0}(kr)+f_{k}K_{0}(kr)
\end{equation}%
where $I_0 $ and $K_0$ are modified Bessel functions of order zero.
The function $I_{0}$ is exponentially divergent for large values of the  argument, then we
choose $f_{I}=0$ in order to have  $f(r)=0$ when $r\rightarrow \infty $. Thus,
$\Xi $ has an oscillatory solution and can be written as

\begin{eqnarray}
\Xi_n(z) &=& \Xi_{(0)n}\cos{k_nz}  \label{xi}
\end{eqnarray}

The energy momentum tensor  that is relevant in the weak field approximation
is $T_{(0)\,\mu \nu }$, which  in this charged model is given by   

\begin{eqnarray}
T_{tt} &=&U\delta (x)\delta (y)+{\frac{Q}{4\pi }}\nabla ^{2}\Big(\ln {\frac{r%
}{r_{0}}}\Big)\noindent  \\
T_{zz} &=&-\tau \delta (x)\delta (y)+{\frac{Q}{4\pi }}\nabla ^{2}\Big(\ln {%
\frac{r}{r_{0}}}\Big) \\
T_{(ij} &=&-Q^{2}\delta _{ij}\delta (x)\delta (y)+{\frac{Q}{2\pi }}\partial
_{i}\partial _{j}\Big(\ln {\frac{r}{r_{0}}}\Big)
\end{eqnarray}%
where, $ U$ and $\tau $ are the energy per unit length and the tension per unit length,
respectively. The other result associated with  the Brane-World parameters
is the polarization of the radiation coming from cosmological distant sources.

The mass $m_{n}$ is small compared with as oscillation frequency of the loop. The relevant part of the 
dilatonic contribution is the $\chi _{n}$ solution which in our model is given by

\begin{eqnarray}
\chi_0 &\!\! \!=& \! \! \!\! 2G_{_{\!\!eff}} C^{^{-1}}\! \!\!\! \! \alpha 
\sqrt{(\nu \!+ \!1)k L}\, e^{-(\nu -1) k L} \Big(\!U\! + \!\tau +\! {\frac{%
Q^2 }{\varepsilon^2 }} \!\Big) \ln(\!{\frac{r }{r_0}}\!)  \nonumber \\
& &  \nonumber \\
\chi_n &\! \! = & \! \! \! 2G_{_{\!eff}}\! C^{^{-1}}\! \!\! \! \alpha \, \,
e^{\nu kL} {\frac{J_{\nu}(x_{n \nu}) }{N_n}} \Big(\!U \!+ \!\tau \!+ \! {%
\frac{Q^2 }{\varepsilon^2 }} \Big)\ln({\frac{r }{r_0}})
\end{eqnarray}

This result is different from the usual case \cite{Bezerra:2004qv} because
in this framework there are ''n" scalar fields corresponding to the oscillation-modes and 
each frequency is related with one energy scale.

\section{Cosmic Microwave Background Radiation and cosmic optical activity}

In this section, we consider the Brans-Dicke parameter and the 
possibility to obtain a model to understand the cosmic optical activity. 
Some authors analyzed the possibility  that optical
polarization of light from quasars and galaxies can  provide evidence 
for cosmical anisotropy \cite{Hutsemekers}. The idea is that in some conditions the
spacetime can  exhibit different properties such as the optical activity
or birefringence [\cite{Das,Nodland97a,Nodland97b}. It is well known that,
as a consequence of vacuum polarization effects in  QED 
\cite{Heisenberg,Weisskopf,Dittrich,Schwinger}, the electromagnetic vacuum
presents a birefringent behavior \cite{Klein,Adler,Brezin,Bialynicka}. For
example, in the presence of an intense static background magnetic field,
there are two different refraction indices depending on the polarization of
the incident electromagnetic wave. This is in fact a very tiny effect which
could be measured in a near future \cite{Heinzl}. The present bounds
for this anisotropy, followed from several experiments\cite{
Semertzidis,Cameron,Zavattini, Zavattini2}, are still far from the expected
value from QED. For this study let us consider the cosmic string background
in the context of the perturbed scalar compactification from 5-dimensions
given by scalar-tensor theory Brane-World scenario, which can be written as

\begin{eqnarray}
\mathcal{S}_{int} & \! \ \! \! \!\! = \! \! \!& \!\int\!\! \! d^4x \sqrt{ \! \! -{\! g}} \Big(-\! \! \! \! \!{\!\!\frac{1}{4}}
  F_{\mu \nu}F^{\mu \nu}\! \!+\! \alpha_+^{^{\! \!(3)}}\! \! \sum_n \! \!\xi_n \partial_{\mu} X_n A_{\nu}
\tilde F^{\mu \nu} \!\Big) \label{actionint2}
\end{eqnarray}
in which we couple the electromagnetic field strenght $F_{\mu \nu }$ and 
the vector potential $A_{\nu }$ with the n modes of the dilaton.  In this context
the equation of the motion for the electromagnetic field becomes

\begin{equation}
\partial_{\mu} F^{\mu \nu} = 2 \alpha_+^{(3)} \sum_n\xi_n \partial_{\mu} X_n
\tilde F^{\mu \nu}  \label{birre}
\end{equation}

The solutions of the Eq.(\ref{birre}) give us the corresponding dispersion
relation

\begin{equation}
(k^{\alpha} k_{\alpha})^2 + (k^{\alpha})k_{\alpha}(v^{\beta} v_{\beta}) =
(k^{\alpha}v_{\alpha})^2 \label{dispersion}
\end{equation}
where $v_{\mu \, n} = \partial_{\mu} X_n$ is responsible for the appearance of the prefered cosmic 
direction, as suggested by the observations \cite{Hutsemekers}. We can expand the dispersion relation 
given by Eq.(\ref{dispersion})in powers of $v_{\alpha}$. In this case we find the dispersion relation 
in power of $\phi_\alpha$, considering the linearized solution in $\phi$ to the first order, giving us

\begin{equation}
k_{n \, \pm }\! = \!\omega_n \pm 2 \alpha_+^{(3)}\alpha_n C^{^{\!-1}}\! \xi^2_n G_{_{\!eff}} \mu \, \,{\hat s} \cos (\gamma )
\end{equation}
with $\mu = (U +
\tau + \Big({\frac{Q }{\epsilon}}\Big)^2)$,  $\omega $ and $k$ being the wave frequency and wave vector,
respectively, the 4-vector $k^{\alpha} = (\omega, k); k = |k|$ and $
\gamma $ the angle between the propagation wavevector k of the radiation
and the unit vector ${\hat s}$.

Let us contextualize our theoretical results in the framework of the
conclusions drawn from the analysis of observational data from quasar emission
performed by  \cite{Nodland97a,Jain,Das}. The angle $\gamma $ between the polarization vector and the galaxies major axis is defined as $<\beta_n >={\frac{r}{2}}\Lambda _{n}\cos(\vec{k},
\vec{s})$, where $<\beta_n >$ represents the mean rotation angle after Faraday
rotation is removed, r is the distance to the galaxy, $\vec{k}$\, the
wavevector of the radiation, and $\vec{s}$ a unit vector defined by the
direction on the sky. The Lorentz breaking imposes that the prefered vector is in $r$ direction. The rotation of the polarization plane is a
consequence of the difference between the propagation velocity of the two modes $ k_{+},k_{-}$, 
the main dynamical quantities. This difference,
defined as the angular gradient with respect to the radial (coordinate)
distance, is expressed as ${\frac{1}{2}}(k_{+}-k_{-})={\frac{d\beta_n }{dr}}$,
where $\beta_n $ measures the specific entire rotation of the polarization
plane, per unit length r, and is given by $\beta_n ={\frac{1}{2}}%
\Lambda _{n}^{-1}r\,\cos \,\gamma $. In the case of the cosmic string solution in
compactified scalar brane world scenarios we have 
\begin{equation}
\Lambda _{n}^{-1}=2\alpha _{+}^{(3)}C^{-1}\alpha_n \,\xi _{n}^{2}G_{_{\! eff}} \, \mu
\end{equation}

If we consider that this result is compatible with the Cosmic Microwave 
Background (CMB) radiation, up to  mode $n$, we obttain 
\begin{equation}
2\alpha_n C^{-1}\xi _{n}G_{_{\!eff}}\, \mu \sim 10^{-6} \label{cmbr}.
\end{equation}

The other important aspect that we must analyse is the hierarchy problem related with the constant $ \alpha_0 $. 
So by compatibility with  CMB radiation, constraint (\ref{cmbr})  must be satisfied. Then,  for $n=0 $ we have 
the zero mode given by 
\begin{equation} 
C^{-1}\alpha_0  \sim  0,5 [12(\nu +1)]^{-1/2}\,e^{-12(\nu-1)} 
\end{equation}
where  $ G_{_{eff}} \, \mu \sim 10^{-6}$ because $G_{_{eff}} \sim 10^{-38}$ and the energy density per unit of lenght $\mu $ of the cosmic string is equal to the symmetry breaking scale $\eta^2$, where $\eta $ is of the order of $ 10^{16}$. 

Let us  analyse the  numerical constraints imposed by CMB radiation and the Brans-Dicke parameter $\omega $ taking into account the compactification modes. 
In this scenario we use the value of $\Lambda^{-1} $ limit of the order of $10^{-32} $ eV \cite{Nodland97a}. 

In order to analyse the contribution of the Lorentz breaking term in the Universe we must understand how the parameters of our model are related with the physical effects. The parameter that is importat here is the dependence of the coupling constant $\Lambda^{-1} $ with the Brans-Dicke parameter $ \omega $ to each modes. This dependence must be a constraint in the range of validity of the  coupling constant $ \Lambda^{-1}$. 

In this framework the zero mode is responsible by an important number of phenomenological implications \cite{Giudice99,Cullen99} and is  only coupled with gravitational fields and, in our approach, with the electromagnetic field strength.  The dependence of the zero mode coupling constant $\Lambda_0^{-1} $ with the 
Brans-Dicke parameter $\omega$ is given by
\begin{equation}
 {\Lambda}_0^{-1}= 10^{-40}[(\nu +1)\,k L]^{1/2}\,e^{(\nu-1) kL},   
\end{equation} 
where we choose the  parameter $ \alpha_+ =10^{-34}$ . We can see from Fig.2  that the  expression of  $ \nu$ as function of $\omega$ is dimension dependent  and the expression is given by $\nu = {(D-1)w+D \over 2(w+1)}$ \cite{Bayona:2008ui}. In our case  D=5 and the  expression reduces to $ \nu = {(4w+5)\over (2w+2)}$ . 
 
\begin{figure}
\centering
\includegraphics[width=8cm]{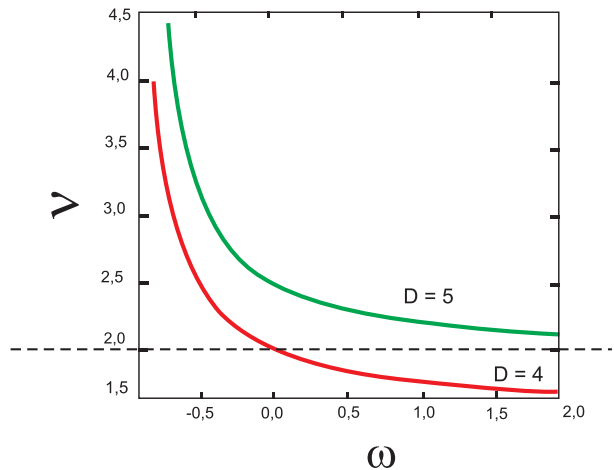}
\parbox{3in}{\caption{\,Dependence of the parameter $\nu$ as a function of $\omega$ corresponding 
to D=4 and D=5. }\label{Figura2}}
\end{figure}

Note in Fig. 2 that when $ \omega \rightarrow \infty $, then $\nu$  goes to 2 and gives us an assymptotical limit $ {\Lambda}_0^{-1} = 0,98 \times 10^{-32}$, if we choose 
$kL = 12$ \cite{Goldberger:1999wh}. This limit corresponds to a lower bound. 
This  model presents another limit where $\omega  $ goes to $ -1$  and  $ {\Lambda}_0^{-1} \rightarrow \infty $. This upper assymptotical  value does not have physical application. We note that the value of $\omega$ taken into account for the fine tune in our model must be linked with experimental data. The range of values that we are considering here is showed in Fig. 3 which presents also  comparison with the massive mode $n=1$.

\begin{figure}
\centering
\includegraphics[width=7.5cm]{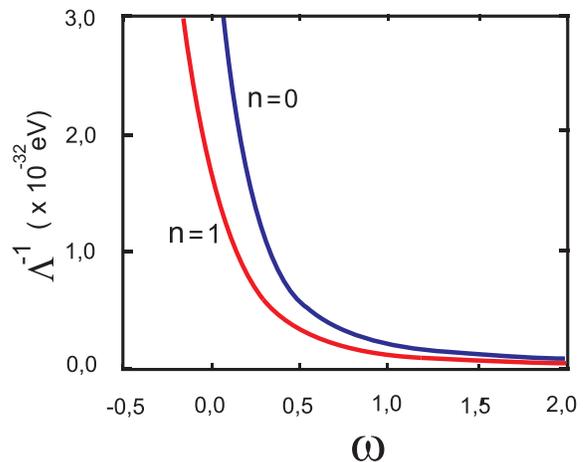}
\parbox{3in}{\caption{Parameter $ \Lambda^{-1} $ for $n\neq 0 $ 
in terms of $\omega$.}\label{Figura3}}
\end{figure}

The behavior of the massive modes is given by the expression

\begin{equation}
\Lambda^{-1}_1 = {10^{-40} \over N_1}J_{\nu}(x_{1 \nu}).
\end{equation}

We can  analyse  the $n=1$ massive mode in  Fig. 3. We note, by comparing with the $n=0$ case that the value of 
 $ \Lambda^{-1} $ decreases. This fact can be understood if we consider that in order to activate this mode we need more energy by the fact that this mode is massive and can be important in events that involves high energy. 

\section{Particle production in the Brane-World Scenario}

In this section we consider the dilaton production given by the oscillating loops of the cosmic string 
that can decay in gauge bosons. 

The loop cosmic string studied in section III couples with the compactified dilaton modes at diferent energies that coincides with the loop oscillator frequencies. During this process,  massive dilatons are copiously emitted with  frequency of oscillation greater than the dilaton mass. 

To take into account the  dilaton decay in gauge bosons we need the interaction with the electromagnetic field. Therefore, we must consider another type of  term where the dilaton with spin-0 particles couples with the electromagnetic field through the mass term with coupling constant, $\alpha_{-}\neq 0$.  

In this section the zero mode is not considered because this mode corresponds to  zero mass. The interaction action responsible for decays into gauge bosons is after
the integration of the $\Omega $-component given by 
\begin{equation}
\mathcal{S}_{int}\!\!=\alpha _{-}\!\!\!\!\sum_{n=1}e^{\nu kL}{\frac{J_{\nu
}(x_{n\nu })}{N_{n}}}\int \!d^{4}x\sqrt{-{g}}X_{n}\!F_{\mu \nu }F^{\mu \nu }\label{actionmass}
\end{equation}

The mass of the dilaton in our model is given by the compactification scales and
the coupling is related with the mode function $\xi _{n}$.

In our model the energy spectrum and angular distribution of the dilaton
radiation can be determined by generalization of the results obtained in \cite{Damour97} to
include the n modes and consider a charged cosmic string.

If we have a loop with length L, $\omega _{n}=4\pi n/L$, the sums are
taken over $n>L/L_{c}$, where $L_{c}=4\pi /m_{n}$. The important point of
our work is that for each n we have a different mass for the dilaton
given by $m_{n} = x_{n\nu}\kappa e^{-\kappa L}$, in accordance with we have  aleady obtained in 
secton II. These masses also have dependence on $ \nu $ that has constraint given by the Brans-Dicke 
parameter as showed in Fig. 4.

The parameters $x_{n \nu}$ can be determined as function of the Brans Dicke parameter $\omega $. The energy range that we consider as a function of $\kappa L $ are specified in Fig. 5, considering 
the constant $ \kappa $ in Planck scale.  
\begin{figure}
\centering
\includegraphics[width=7.5cm]{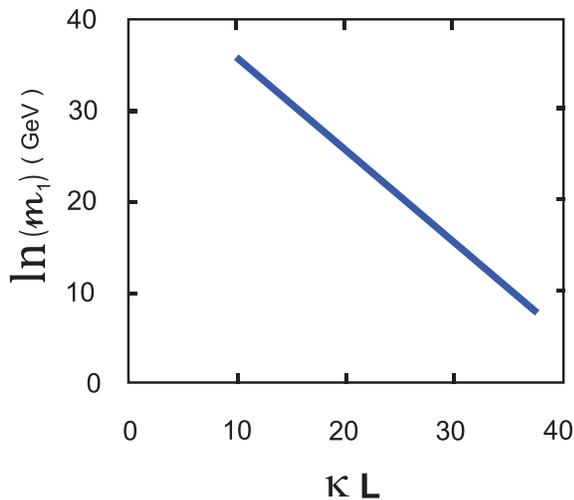}
\parbox{3in}{\caption{\, Mass for  n=1 as a function  of $\kappa L$.
 }}\label{Figura4}
\end{figure}
In the context of the dilaton, the energy scales are compatible with the references \cite{Ellis86}.
In our model, in order to find masses with this energy of the order of TeV for $n= 1$, we have $ k L \sim 40 $. 
In this process there is the possibility to generate light gauge bosons.
However if we consider  $\kappa L =12$, as in last section, the particles masses are of the order of 
$(\sim 10^{14}) GeV$. These values for the masses can be mensurable in the Pierre Auger experiment, for example.  The decays of these massive particles can be detected, in principle, with lower energies.

The energy and particle radiation rates when the loop is such that $L>>L_{c}$ have a mode
dependence which can be represented as
\begin{eqnarray}
\dot E_{X_n} &=& \Gamma_{X_n} \alpha_n^2 G_{eff} U^2 (L/L_c)^{-1/3} \\
\dot N_{X_n}& = &\tilde\Gamma_{X_n} \alpha_n^2 G_{eff} U^2
m_{X_n}^{-1}(L/L_c)^{-1/3}
\end{eqnarray}
The resulting constraints are relevant if we consider the life-time of the
dilaton, that are determined by the mass and couplings. The
corresponding lifetime is
\begin{equation}
\tau_n = {\frac{4M_p^2 }{N_F \tilde \alpha^2_{F_n} m^3_n}}
\end{equation}
where $\tilde \alpha_{F_n} = \alpha_-\alpha_n C^{-1} \xi_n$ and
$N_F$ is the number of gauge bosons with masses $<< m_{X_n} $.
In Fig. 5  we consider $m_{X_n} \sim 1 $
TeV. In this case all standard-model gauge bosons should be included ($N_F = 12$). We note that the lifetime 
of the dilatons has a fine structure constant given by the Brans-Dicke parameter. It is important to ajust 
with the experimental data. Here we also have the constraints given by the Brans-Dicke parameter $\omega $. For $\omega = -1 $ the lifetime agrees at infinity, giving us the possibility that massive particles decay in the primordial Universe and probably could be measured in the detectors nowadays.

\begin{figure}
%\vspace{1 true cm}
\centering
\includegraphics[width=8cm]{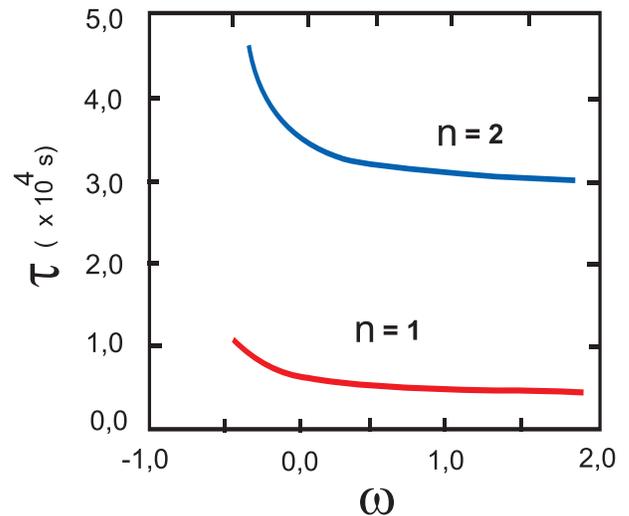}
\parbox{3in}{\caption{This graphic represent the the lifetime of the particles in function of the Brans-Dicke parameter.
 }}\label{Figura5}
\end{figure}

\section{Conclusions}

It is possible to construct a Brane-World scenario including gravity in which we can analyse 
the effects of the Lorentz Breaking in the framework of a cosmic string configuration. With this 
aim we can assume that the scalar tensor theory is realized in 5-dimensions. In this scenario it
is showed that the cosmological birefringence is connected with the Brans-Dicke parameter which
is constrained by the date of the Cosmic Microwave Background Radiation and by the compactification 
modes. The limits of the rotation angle depends on the Brans-Dicke parameter and decreases with the
increasing of masses of the correpsonding massive modes. 

In this framework of saclar-tensor theory of gravity in 5-dimensions, the ineraction
terms can be taken with and without parity violation. These terms are connected
with the cosmic string configuration we are considering. The Lorentz breaking term
is responsible for generation of charge and the part that does not present the parity
violation put limits on the mass scale of the particles which decay into
light dilaton. The interaction Lagrangian (\ref{actionmass}) is responsible for the decay into light gauge 
bosons, where are the sources of cosmic strings. These depend on the modes given by the 5-dimensional 
Brane-World scenario.

\vspace{1 true cm}

{\large \textbf{Acknowledgments:}} The authors would like to thank
(CNPq-Brazil) for financial support. C. N. Ferreira also thank Centro
Brasileiro de Pesquisas F\'{\i}sicas (CBPF) for hospitality. V. B. Bezerra 
also thanks FAPESQ-PB (PRONEX) and FAPES-ES (PRONEX). G. de A. M. tanks 
FAPESQ - PB (PRONEX).

\end{document}